\documentclass[preprint,12pt,3p]{elsarticle}

\usepackage[utf8]{inputenc}
\usepackage[T1]{fontenc}
\usepackage{textcomp}
\usepackage{amssymb}
\usepackage{float}
\usepackage{subfigure}
\usepackage{amsmath}
\usepackage{cleveref}
\usepackage{bm}
\usepackage{upgreek}
\usepackage{bigints}
\usepackage[nomarkers]{endfloat}

\biboptions{numbers,sort&compress}

\journal{Journal of Computational Science}

\graphicspath{{figures/}}

\begin{document}

\begin{frontmatter}

\title{Large-scale multiscale modeling of phase transformation in nanocrystalline materials: Atomistic and Phase-Field methods}

\cortext[cor1]{Corresponding author: Mehrdad Yousefi
Department of Materials Science and Engineering, Clemson University, 161 Sirrine Hall, Clemson, SC 29634, USA
Email: yousefi@clemson.edu Phone: 1-864-624-2839}

\author[cor1]{Mehrdad Yousefi\corref{cor1}}
\ead{yousefi@clemson.edu}

\begin{abstract}
In this research, atomistic molecular dynamics simulations are combined with mesoscopic phase-field computational methods in order to investigate phase-transformation in polycrystalline Aluminum microstructure.
In fact, microstructural computational modeling of engineering materials could help to optimize their mechanical properties for industrial applications (e.g. directional solidification for turbine blades).
As a result, a multiscale modeling approach is developed to find a relation between manufacturing variables (e.g. temperature) and microstructural properties of crystalline materials (e.g. grain size), which could be used to develop an advanced manufacturing process for sensitive applications.
The results show that atomistic modeling of grain growth could be used as a first-principle approach in order to study phase transformation's kinetics, which could capture morphology of polycrystalline materials more accurately.
On the other hand, phase-field mesoscopic approach needs less computational efforts, but still it relies on semi-empirical data to capture accurate phase transformation regimes, which makes this approach suitable for rapid examining of new manufacturing conditions as well as its effects on microstructural properties of polycrystalline materials.
\end{abstract}


\end{frontmatter}


\section{Introduction}
\label{sec1}

The nanocrystalline microstructure of inorganic materials (\textit{e.g.}, metals) could directly affect their engineering properties. Even though the grain size of polycrystalline materials can be controlled directly by time and temperature of the annealing process, the mechanism of grain growth is not fully understood \cite{MD3}. Mesoscopic computational methods can be used to study the thermodynamics of grain growth, extract the chemical potential functional of the system, and solve a non-Fickian diffusion equation (\textit{e.g.}, Allen-Cahn or Cahn-Hilliard equations) to obtain the spatial distribution of a multiphase system \cite{Phase-Field, Phase-Field-1, Phase-Field-2}.

Recently, due to the rapid growth of high-performance computing capabilities, it is possible to approach a multiphase system by using atomistic computational methods \cite{MD1, MD2}. This work aims to find a relation between microstructural configurations or morphology of crystalline materials with the engineering controlling variables (\textit{e.g.}, time and temperature). Understanding the microstructural characteristics under varying conditions could help engineers to design polycrystalline materials for applications where a specific morphology is preferable (\textit{e.g.}, turbine blades) \cite{turbine1,turbine2}. In this project, both mesoscopic and atomistic approaches were taken to examine the advantages and limitations of each. The effects of temperature and initial morphology on the final microstructure is studied to find a relation between engineering variables (\textit{e.g.}, temperature) and configuration of the polycrystalline material.

\section{Material and methods}
\label{sec2}

The molecular dynamic simulations were carried out using the LAMMPS (large-scale atomic/molecular massively parallel simulator) software \cite{LAMMPS} with an Embedded Atom Method (EAM) potential \cite{EAM1,EAM2} to elucidate the effects on the grain boundaries (GB) when the number of grains and temperature are increased for a pure aluminum (Al) crystal. The EAM potential can be described by eq. (\ref{eq:1}), where $E_{i}$ is the total energy of atom i, $F_{\alpha}$ is the surrounding energy function, $\rho_{\alpha}$ is the electron density and $\phi_{\alpha \beta}$ is the double-well interaction potential of $\alpha$ and $\beta$ type atoms.\\
\begin{equation} \label{eq:1}
 E_{i} = F_{\alpha} (\sum_{j \neq i} \rho_{\alpha} (r_{ij})) + \frac{1}{2} \phi_{\alpha \beta} (r_{ij})
\end{equation}
The EAM functional form, which is derived from \cite{EAM3, EAM4}, is used for the aluminum pair-wise interaction for all molecular dynamic simulations.

\begin{figure}[H]
 \centering
 \includegraphics[width={0.8\textwidth},keepaspectratio]{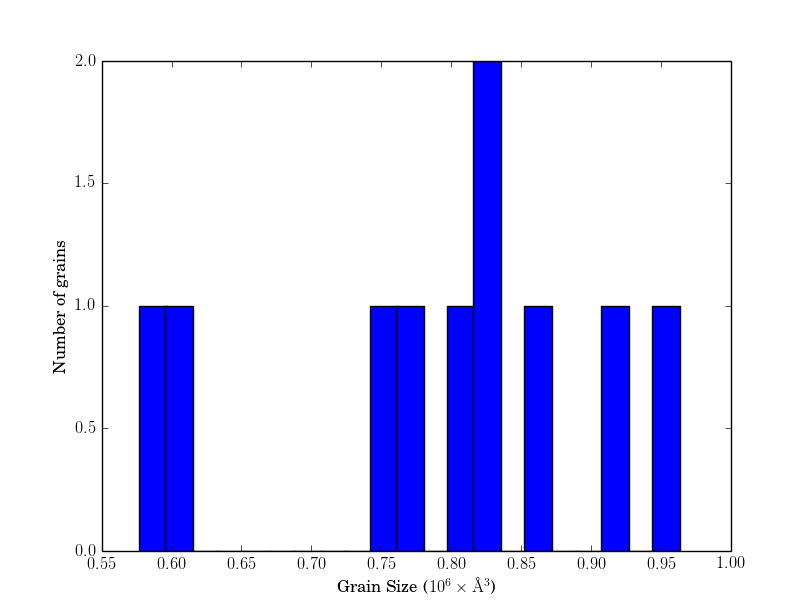}
 \caption{Grain size distribution for 10 grains randomly oriented; generated using Atomsk software.}
 \label{fig:1}
\end{figure}

Four 200x200x200 (the length scale is \AA) aluminum polycrystalline structures with 5, 7, 9, or 10 grains were constructed using the Atomsk software \cite{atomsk}, with a 4.046 \AA{} minimum distance between the aluminum atoms. The grains were oriented randomly and the size is distributed along a normal distribution (Figure \ref{fig:1}). The timestep and final run time used for the molecular dynamic simulations were 0.005 ps and 5000 ps, respectively. The simulation box contained 483,115 atoms for all simulations. The NVT ensemble is used to model grain growth and periodic boundary conditions were imposed on the boundaries of the simulation box to mimic an infinite computational domain. Each polycrystalline structure is studied under three different temperatures (\textit{e.g.}, 500K, 600K and 700K) to examine the effect of temperature on the final microstructural morphology of the system. 
Furthermore, in order to show the applicability of our computational framework to study more complex systems, particularly multi-components and multi-phases systems, one simulation of Al-Cu system with 10 grains (5 Al grains and 5 Cu grains, 50\%/50\% atomic chemical composition of nanocrystalline alloy) is constructed at 600K for 5000 ps with periodic boundary condition and using same NVT ensemble to study microstructral evolution in this nanocrystalline alloy. In order to model the interaction of Copper-Copper and Copper-Aluminum atoms, an alloy EAM potential, which is constructed based on \cite{ZHOU2016752,apostol2011,LIU19993227}, is added to previous framework for pure Al nanocrystalline materials simulations. The aim is that to show applicability and convenience extension of our multi-scale computational framework to study complex engineering systems.

Due to researches of \cite{doi:10.1021/acs.jpcc.6b10908} and \cite{ZHOU2016752}, where the EAM potentials for Al-Al and Al-Cu interactions are extracted respectively, the developed potentials to capture physical properties of nanocrystalline materials such as their melting point, lattice constant, module of elasticity, etc. are valid in the range of phase-transformation temperatures in this study (500K to 700K) and could capture inter-atomic interactions with reasonable accuracy. We refer readers to the mentioned references (\cite{doi:10.1021/acs.jpcc.6b10908} and \cite{ZHOU2016752}) for more detail information about the construction of these force-fields for Al-Al and Al-Cu interactions.

For the mesoscopic simulations, the continuum-field model \cite{Phase-Field,Phase-Field-1, Phase-Field-2} is used to extract the time evolution of grains under certain kinetically controlled conditions. The Gibbs free energy functional plays a critical role in continuum-field models, as described by the Ginzburg-Landau form (eq. (\ref{eq:2})).
\begin{equation} \label{eq:2}
 F = \int_{\Omega} [f_{local}(\eta_{i} (\vec{r})) + \frac{1}{2} \kappa \sum_{i} \nabla^{2} \eta_{i} (\vec{r})] d^{3} \vec{r}
\end{equation}
The first term in eq. (\ref{eq:2}) is related to the volumetric contribution and the second term, which is proportional to the gradient of curvature, is the surface curvature contribution. Since the surface contribution is much higher than the volumetric contribution, the system will try to minimize its energy functional by growing the grain boundaries which decreases the surface energy. Using the Euler-Lagrange extremum approach, the governing equation could be defined by,
\begin{equation} \label{eq:3}
 \frac{\partial \eta_{i}}{\partial t} = -L (\frac{\delta F}{\delta \eta_{i}}) = -L (\frac{\partial f_{local}}{\partial \eta_{i}} - \kappa \nabla^{2} \eta_{i}).
\end{equation}
The well known double-well potential \cite{Phase-Field,Phase-Field-1,Phase-Field-2} is used to model $f_{local}$, as defined by eq. (\ref{eq:4}).
\begin{equation} \label{eq:4}
 f_{local} = -\frac{\alpha}{2} \sum_{i} \eta_{i}^{2} + \frac{\beta}{4} (\sum_{i} \eta_{i}^{2})^{2} + (\gamma-\frac{\beta}{2}) \sum_{i} \sum_{j \neq i} \eta_{i}^{2} \eta_{j}^{2}
\end{equation}

The dimensionless values used for all mesoscopic simulations are: $\alpha = 1.0$, $\beta = 1.0$, $\gamma = 1.0$, $\kappa = 2.0$ and $L = 1.0$. The initial polycrystalline microstructures are generated using Dream3D software \cite{Dream3D}. Periodic boundary conditions were imposed on all the surfaces of the cubic simulation box (200x200x200 \AA). Equation (\ref{eq:3}) is discretized by using a Python script of the Crank-Nicolson finite difference method, primarily developed by the authors. 

\begin{figure}[H]
 \centering
 \includegraphics[width={0.8\textwidth},keepaspectratio]{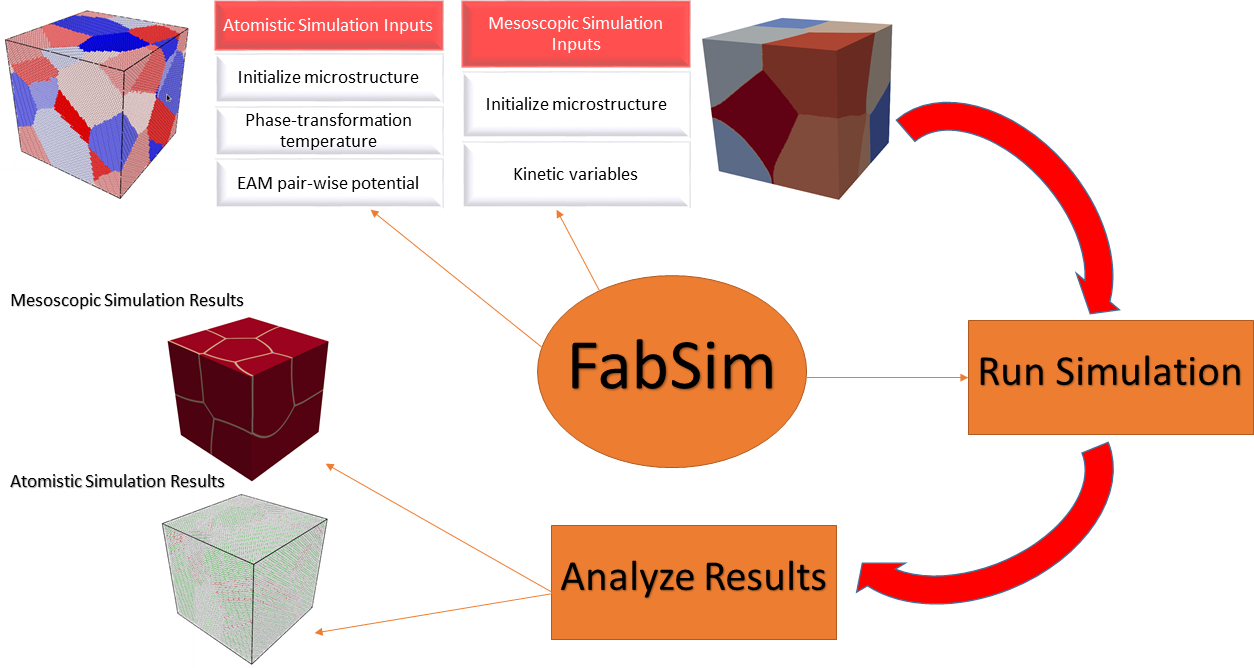}
 \caption{Automatic pipline to construct atomsitc (molecular dynamics) and mesoscopic (phase-field) simulations based on given initial microstructure, phase-transformation temperature, and atomistic pair-wise potential (in case of molecular dynamics simulations).}
 \label{fig:ensemble}
\end{figure}

At the end all the mentioned steps are constructed based on automatic computational ensemble (i.e. FabSim) \cite{GROEN2016375,doi:10.1021/ci100007w,doi:10.1021/acs.nanolett.5b03547} to construct atomistic (i.e. molecular dynamics) and mesoscale (i.e. phase-field) simulations to ensure all the steps in the modeling pipeline, from initializing the initial microstructure based on user-defined number of grains, phase-transformation temperature, and atomistic potential for molecular dynamics simulations (i.e. in our case EAM pair-wise potential) would be done with minimum user interaction in order to facilitate the use-ability of the proposed framework to extend the materials science research for investigating advanced nanocrystalline materials. The detail of this automatic computational ensemble for both molecular dynamics and phase-field simulations are shown in Figure \ref{fig:ensemble}. Due to Figure \ref{fig:ensemble}, the initial microstructure of atomistic and mesoscopic simulations will be generated by using Atomsk and Dream3D softwares, then phase-transformation temperature and pair-wise potential (i.e. EAM potential) will be chosen based on chemical composition of initial microstructure, and finally these automatically generated microstructure and corresponding potentials will be forwarded to atomistic or phase-field solvers and then final results and analysis will be extracted by using Python bindings in Ovito software \cite{ovito}.

In the next section, the results of both mesoscopic and molecular dynamic simulations will be discussed in terms of their morphology, advantages and limitations, and reliability to deploy them for real industrial cases.

\section{Results}
\label{sec3}

\subsection{Temperature effects}
The results for the polycrystalline structures at different temperatures are visualized using Ovito visualization software \cite{ovito}. The visualization software can identify the types of crystal unit cell structures and the grain boundaries. During the grain growth process, volumetric and surface Gibbs free energies compete against each other to dominate the grain growth mechanism. Since the surface contribution is much higher than the volumetric contribution, the system will try to minimize its energy functional to reduce the surface energy. Surface diffusion might dominate the grain growth as the temperature increases due to the enhanced mobility of atoms.

\begin{figure}[H]
 \centering
 \includegraphics[width={0.8\textwidth},keepaspectratio]{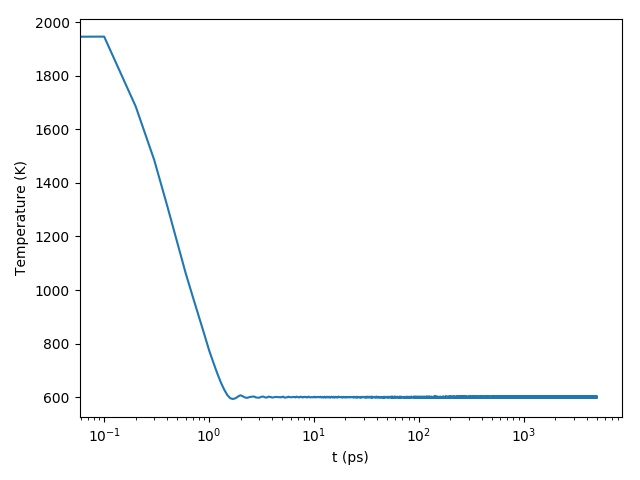}
 \caption{Phase-transformation temperature for Aluminum nanocrystalline material at 600K with 10 initial grains.}
 \label{fig:tempandenergy1}
\end{figure}

\begin{figure}[H]
 \centering
 \includegraphics[width={0.8\textwidth},keepaspectratio]{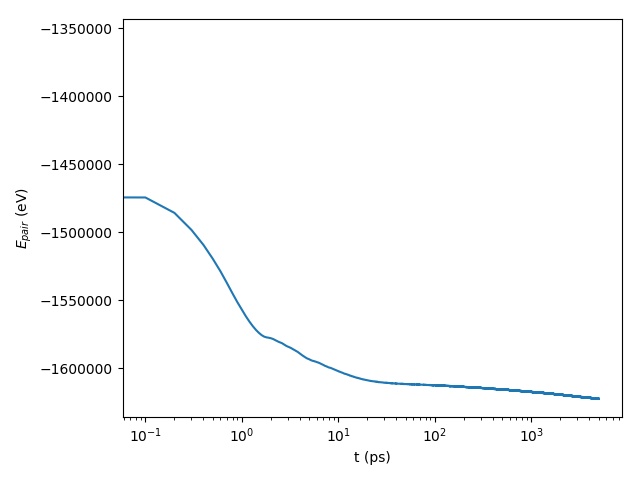}
 \caption{Pair-wise energy of the system for Aluminum nanocrystalline material at 600K with 10 initial grains.}
 \label{fig:tempandenergy2}
\end{figure}

\begin{figure}[H]
 \centering
 \includegraphics[width={0.8\textwidth},keepaspectratio]{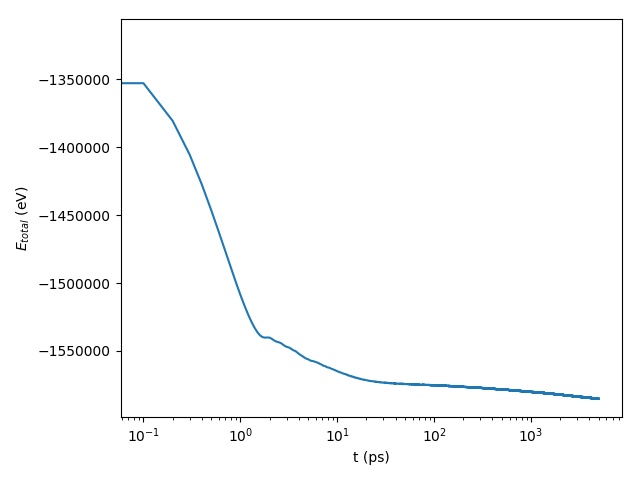}
 \caption{Total energy of the system for Aluminum nanocrystalline material at 600K with 10 initial grains.}
 \label{fig:tempandenergy3}
\end{figure}

The observed phase transformation in this Al nanocrystalline system, despite short time-scale of this study (5000 ps), is attributed to higher surface energy of this nanocrystalline materials because of nano size grain boundaries in comparison to bulk materials, which resulted in much lower activation energy for this Al nanocrystalline microstructure. This findings is consistent with several other researches that studies phase-transformation in nanocrystalline materials, specifically grain growth of metallic materials, that short phase-transformation time-scale observed predominantly \cite{MD3,KO201790,doi:10.1063/1.5044792}. As a result, in order to show precise convergence of phase-transformation in our simulation, the evolution of phase-transformation temperature, pair-wise energy, and total energy of the system are plotted in Figures \ref{fig:tempandenergy1}, \ref{fig:tempandenergy2}, and \ref{fig:tempandenergy3} respectively for the simulation of Aluminum with 10 grains at 600K. Due to Figures \ref{fig:tempandenergy1}, \ref{fig:tempandenergy2}, and \ref{fig:tempandenergy3}, phase-transformation temperature, pair-wise energy, and total energy reached a plateau which shows that phase-transformation converged to its physical state, which is consistent with several findings from \cite{HASLAM2001293,MIYOSHI2018118,Okita2016ISIJINT-2016-408,C8CE00767E,foiles2012}.

\begin{figure}[H]
 \centering
 \includegraphics[width={0.8\textwidth},keepaspectratio]{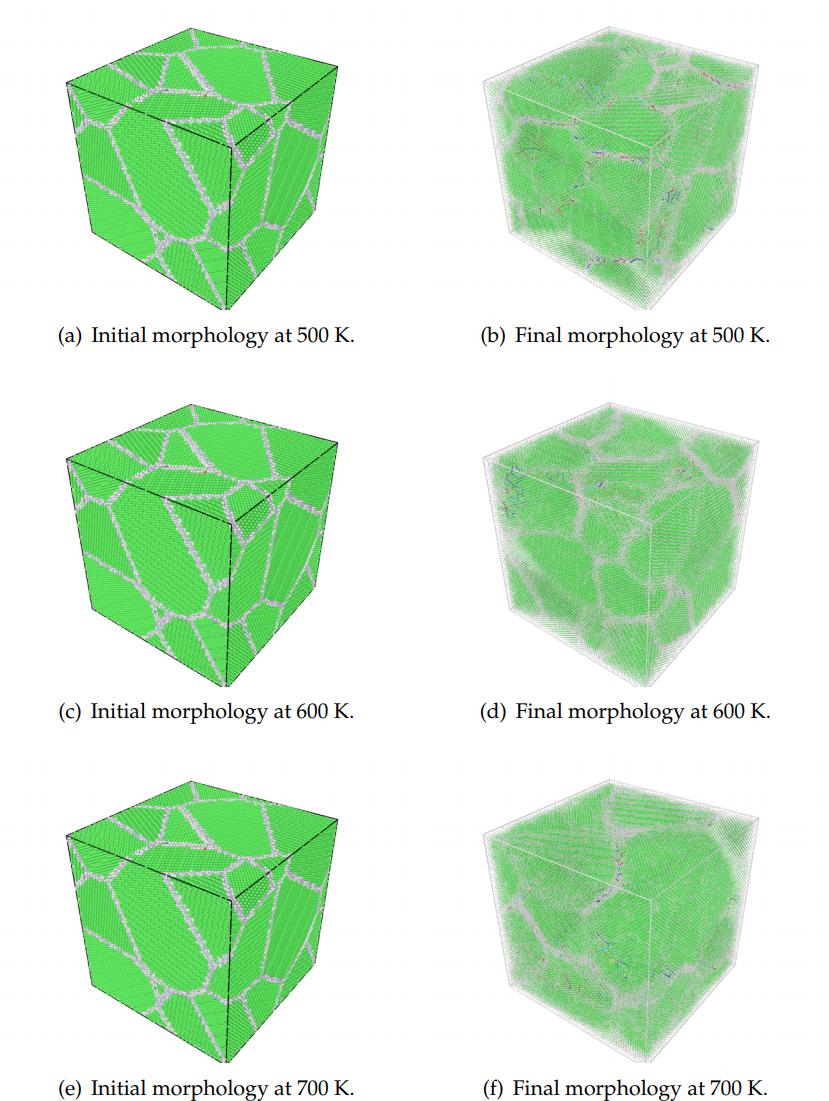}
 \caption{Aluminum GB structures with 9 grains at three different temperatures.}
 \label{fig:2}
\end{figure}

The initial configuration of the polycrystalline structures (Figure \ref{fig:2}) has four or higher order grain junctions. These higher order grain junctions are not favorable regarding the surface Gibbs free energy. Considering the melting point of aluminum is 933.5 K, the temperatures investigated in this work are lower, impacting the mobility of the atoms in the system. After 5000 ps, the higher order junctions were not eliminated from the 9 grain polycrystalline structure at 500 K (Figure \ref{fig:2}) but the number of three grain junctions increased because the surface energy is minimizing its value across the grain boundaries. Diffusion mechanisms (\textit{e.g.}, creep) could be activated above 0.5$T_{m}$, where $T_{m}$ is the absolute melting point of the material \cite{creep}. Indicating at higher temperatures the increased mobility of the atoms in the system can decrease the grain boundary surfaces. For the 9 grain polycrystalline structure at 600 K (Figure \ref{fig:2}), the higher order grain junctions were eliminated in the final structure. However, at 700 K there is an abnormal grain growth in the final structure due to the lack of higher order junctions in the initial structure. The final 9 grain polycrystalline structure at 600 K is suitable for industrial applications because only three grain junctions exist in the system. A material dominated by three grain junctions is an important characteristic because the three grain junctions balance out the surface tension allowing the microstructure to be isotropic, which implies the mechanical properties of the system are isotropic. The abnormal grain growth (Figure \ref{fig:2}) for the 9 grain polycrystalline structure at 700 K is not desirable because large grains can reduce the mechanical properties of a material significantly.

One of the mechanisms in plastic deformation of polycrystalline materials is dislocations \cite{dislocation}. Dislocations can be interpreted as a missing crystal plane which could facilitate the plastic deformation of polycrystalline materials. If the density of dislocations is high, the dislocations will be fixed, and further movement or distortion is not possible. However, the mechanism of dislocation formation is not fully understood because of the complex morphology in polycrystalline materials. One of the features of Ovito is an algorithm that can identify the dislocations or missing planes in the crystalline structure. According to Figure \ref{fig:3}, the grain boundaries are the source of dislocations, as shown by the red and blue lines. Increasing the temperature from 500 K to 600 K corresponds to an increased number of dislocations (cref. Figure \ref{fig:3}), but at 700 K, the mobility of the aluminum atoms eliminates all the grain boundaries and dislocations which could make the polycrystalline material ductile. Regarding engineering applications, strength is essential (\textit{e.g.}, ductility) and the 9 grain polycrystalline structure at 600 K has the most favorable microstructure because the final system is composed solely of three grain junctions and the density of the dislocations in the system is minimal. In the next section, the results for the mesoscopic simulations will be discussed and compared to the molecular dynamic simulation results.

\subsection{Multi-components nanocrystalline simulation of Al-Cu alloy}

\begin{figure}[H]
 \centering
 \includegraphics[width={0.8\textwidth},keepaspectratio]{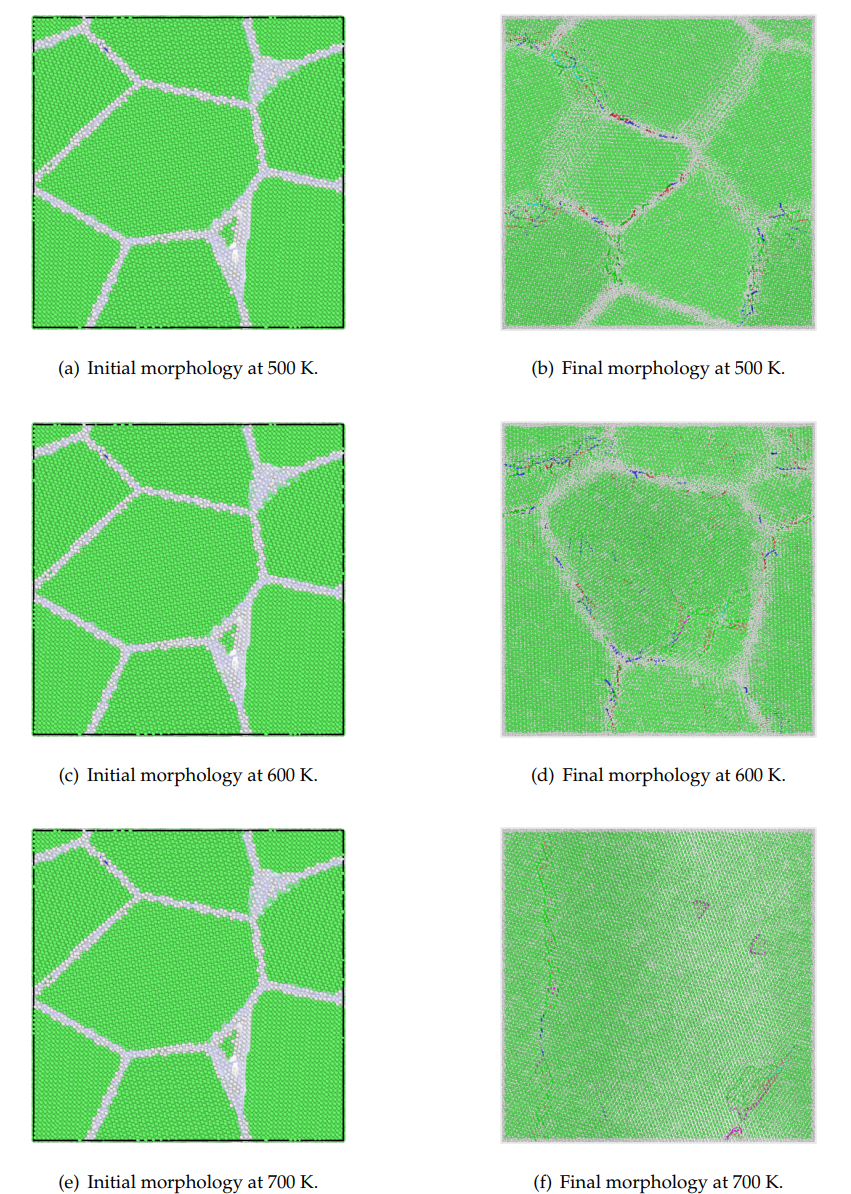}
 \caption{2D section of the aluminum GB structures with 7 grains at three different temperatures.}
 \label{fig:3}
\end{figure}

\begin{figure}[H]
 \centering
 \includegraphics[width={0.8\textwidth},keepaspectratio]{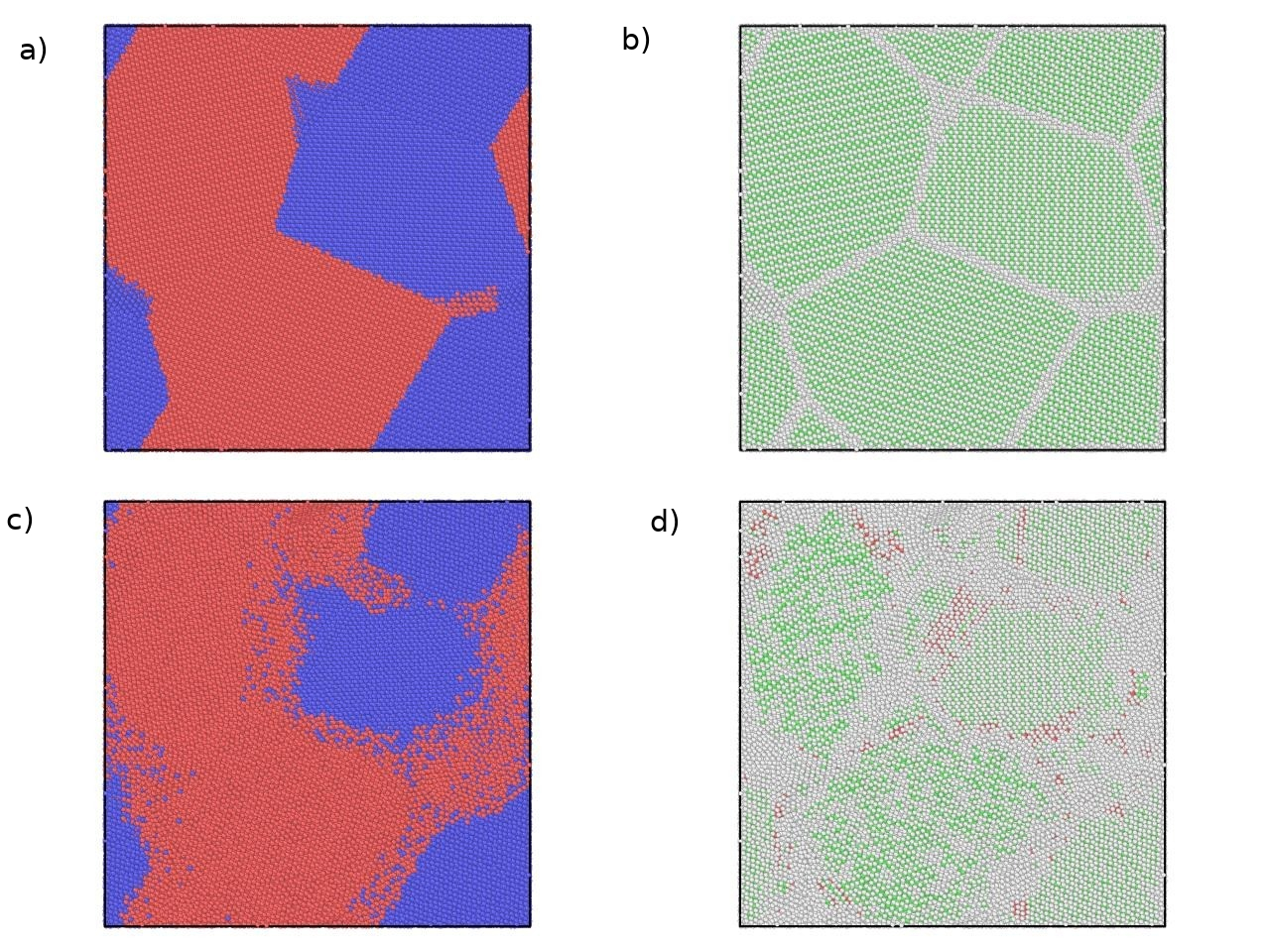}
 \caption{2D section of the Al-Cu GB structures as well as their coloring based on atom types (blue: Al, red: Cu) with 10 grains at 600K, a) initial microstructure colored by atom types, b) initial microstructure colored by grain boundaries, c) final microstructure colored by atom types, d) final microstructure colored by grain boundaries.}
 \label{fig:alcu}
\end{figure}

In order to show the applicability of the proposed computational framework to study metallic nanocrystalline materials, a multi-components nanocrystalline structure of Al-Cu alloy (50\%/50\% Al-Cu atomic percentage chemical composition) with 10 initial grains at 600K is investigated and initial and final microstructures are analyzed (cref. Figure \ref{fig:alcu}). Due to Figure \ref{fig:alcu}, it could be understood that presence of high melting point component (i.e. Copper) could prevent abnormal growth observed in 600K of pure Aluminum (cref. Figure \ref{fig:3}), which is consistent with previous research from \cite{Verestek2017}. Furthermore, the final structure of this nanocrystalline Al-Cu alloy (cref. Figure \ref{fig:alcu}d) shows that there is an increase in grain boundary regions in comparison to pure Aluminum at 600K (cref. Figure \ref{fig:3}), which are identified as the  source of dislocation and would help to increase the mechanical strength of Al-Cu nanocrystalline alloy. Also, due to Figure \ref{fig:alcu} and Figure \ref{fig:3}, it could be understood that the presence of high melting component with lower diffusion coefficient limited the mobility of low melting component (i.e. Al) and resulted in a lower grain size at the final stage of phase-transformation, which again would help to increase mechanical strength of this nanocrystalline structure based on Hall-Petch relation. 


\subsection{Mesoscopic simulations}
The mesoscopic simulations investigated in this work, can be called an isothermal continuum-field model. Meaning, no temperature gradient exists in the model. Therefore, a heat transfer equation needs to be coupled to a continuum-field equation to account for the temperature gradient. In addition, the semi-empirical variables used in the continuum-field model (\textit{i.e.}, $\alpha$, $\beta$, $\gamma$, $\kappa$ and $L$) should be considered as temperature-dependent kinetic parameters instead of constant values. Unfortunately, extracting the kinetic values for polycrystalline materials is not easy and several microscopic experiments of polycrystalline microstructures are needed to find the relationship between time and temperature. In this particular case, molecular dynamics could be assumed as a first principles approach to the grain growth problem because the only thing that needs to be specified in a molecular dynamic simulation is the pair-wise interaction, which is fully known for metallic materials. This could be considered as an advantage over mesoscopic simulation which suffers from lack of semi-empirical data.

\begin{figure}[H]
 \centering
 \includegraphics[width={0.8\textwidth},keepaspectratio]{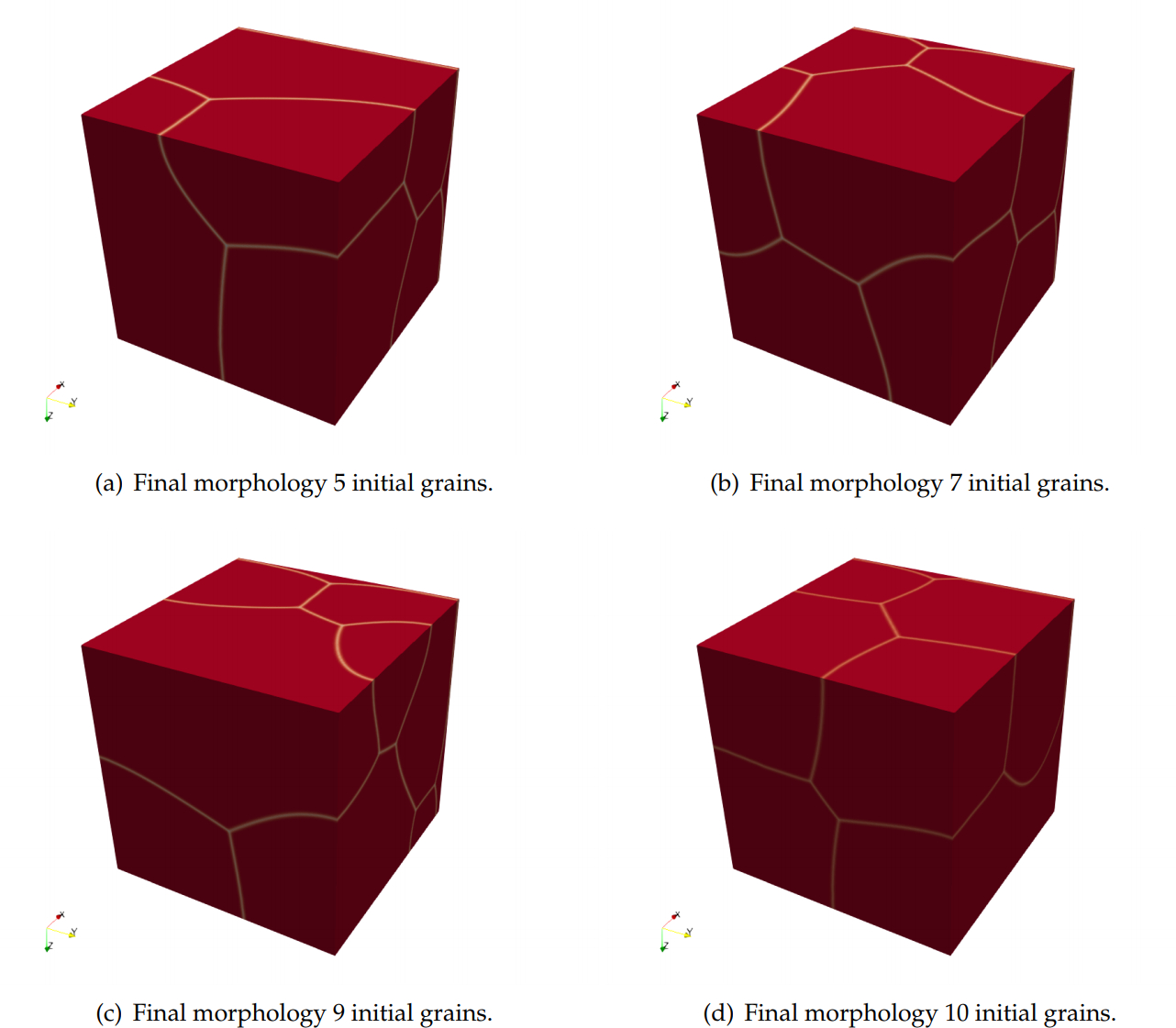}
 \caption{Final microstructures of the mesoscopic simulations.}
 \label{fig:4}
\end{figure}

The final microstructures of the mesoscopic simulations are shown in Figure \ref{fig:4} for four different initial grain structures. In comparison to the molecular dynamics model, the mesoscopic simulations do not give the types of crystal unit cell structures. Also, the grain boundary width (\textit{i.e.}, $\kappa$) is taken into account as a parameter which defines the width of the diffusion region or grain boundaries. Referencing Figure \ref{fig:4}, the initial finer microstructure (10 initial grains) could be identified as the reliable microstructure in terms of grain size and three grain junction points. It appears, because the mesoscopic model is isothermal and the kinetic parameters are constant values, mesoscopic simulations cannot predict the observed abnormal grain growth seen in the molecular dynamic simulations (cref. Figures \ref{fig:2} and \ref{fig:3}).

\subsection{Atomistic and mesoscopic simulations: grain size comparison}

\begin{figure}[H]
 \centering
 \includegraphics[width={0.8\textwidth},keepaspectratio]{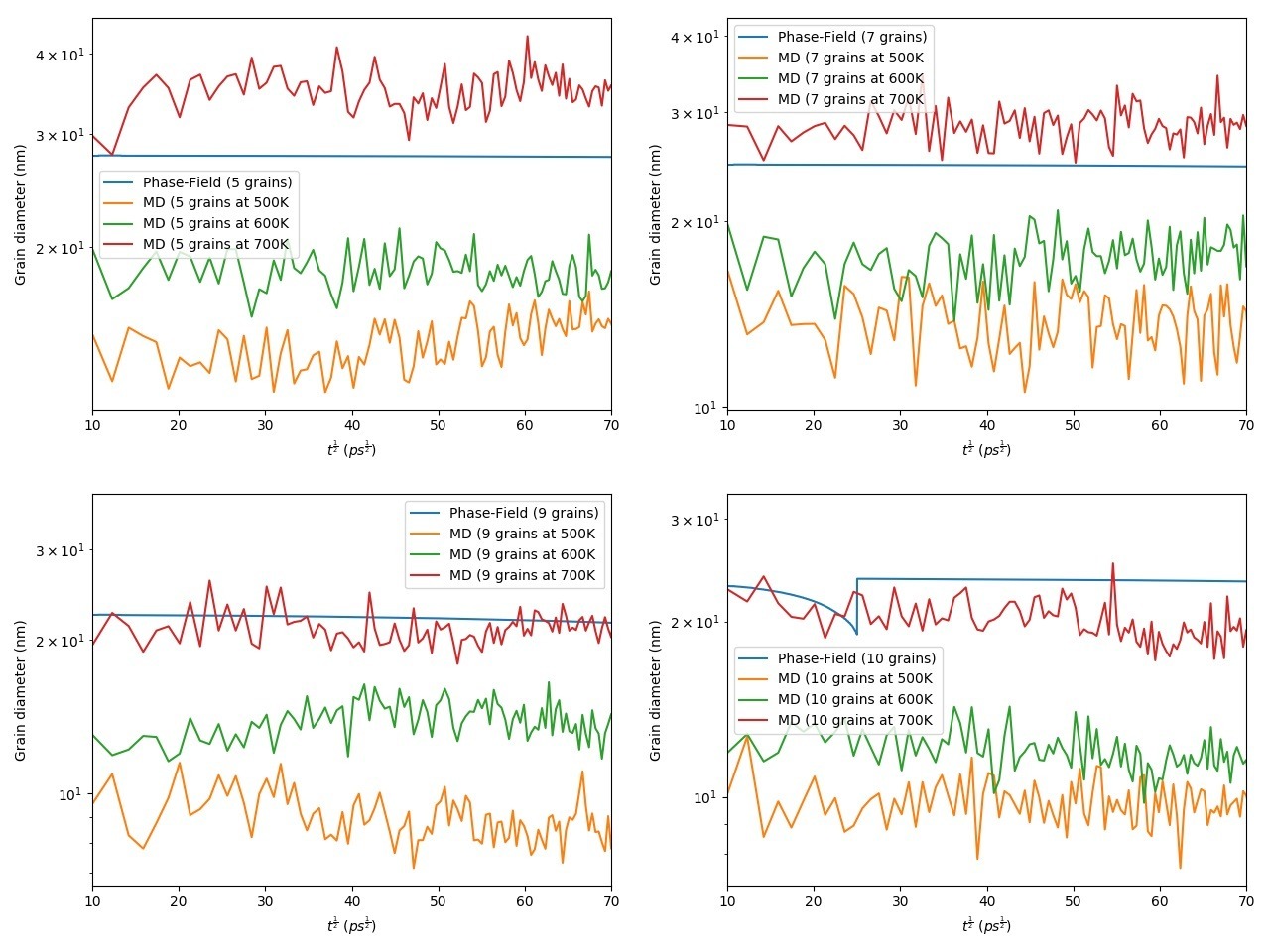}
 \caption{Grain sizes plotted versus square root of time for different microstructures at different phase-transformation temperatures for both molecular dynamics and phase-field simulations.}
 \label{fig:grainsizes}
\end{figure}

In order to compare different microstructures at different phase-transformation temperatures quantitatively for atomistic (MD) and mesoscopic (phase-field) approaches, the grain diameter versus square root of phase-transformation time is plotted in Figure \ref{fig:grainsizes}. Note that the diagrams in Figure \ref{fig:grainsizes} are plotted after 100 ps due to reaching convergence plateu of temperature, pair-wise energy, and total energy according to Figures \ref{fig:tempandenergy1} to \ref{fig:tempandenergy3}.  Due to Figure \ref{fig:grainsizes}, it could be understood that phase-field simulation result of grain size is close to grain size evolution of atomistic simulations at 700K for all the initial microstructure configurations. In fact, this approach could be used to extract kinetic variables of phase-field (mesoscopic) approach at different temperature by matching the grain diameter evolution with molecular dynamics results. The most accurate agreement between molecular dynamics and phase-field simulations is observed at grain diameter evolution diagram of initial 9 grains structure at 700K, which means the kinetic variables that are extracted from \cite{Phase-Field,Phase-Field-1, Phase-Field-2} could capture phase-transformation behavior of pure Aluminum at 700K precisely. At the end, it could be noted that based on findings from Figure \ref{fig:grainsizes}, it could be inferred that molecular dynamics propose much more flexible computational framework, which could capture different phase-transformation regimes of nanocrystalline materials more precisely \cite{Holm1138}.

\subsection{Figures}
\begin{figure}[H]
 \centering
 \includegraphics[width={0.8\textwidth},keepaspectratio]{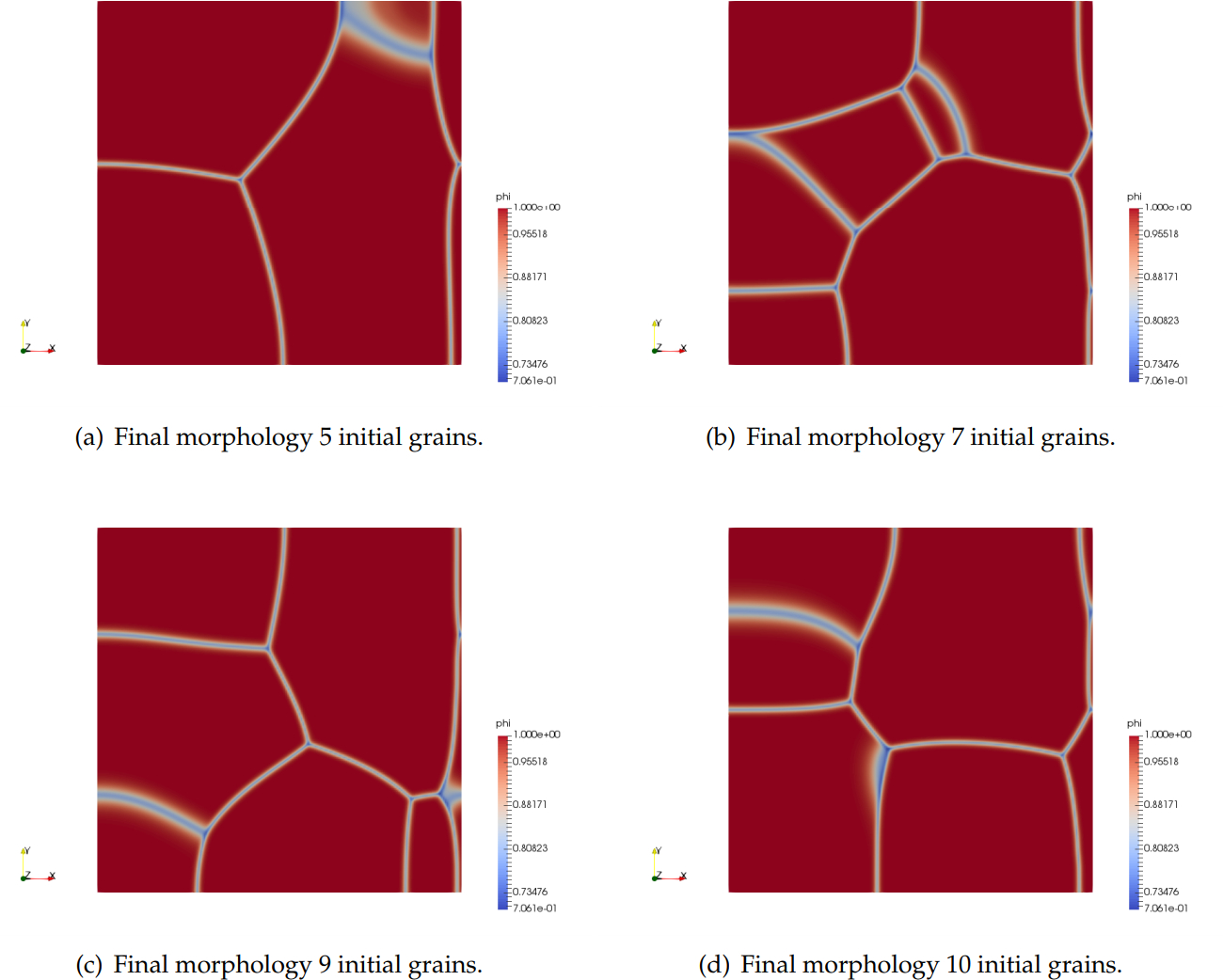}
 \caption{2D section of the final microstructures for the mesoscopic simulations.}
 \label{fig:5}
\end{figure}

\section{Conclusions}
\label{sec4}

In this research, the grain growth in polycrystalline materials are studied by two different approaches, molecular dynamics and mesoscopic models. According to the results, the mesoscopic simulations are faster than the molecular dynamic simulations. If a study needed to consider a larger spatio-temporal scale, mesoscopic models could be a better fit than molecular dynamics. However, the mesoscopic computational model in this work suffers from the lack of semi-empirical data for the kinetic variables and those values are not easy to extract them for real polycrystalline materials. On the other hand, molecular dynamics could tackle this problem as a first principles approach. Molecular dynamics could predict the effect of temperature and initial grain size on final microstructure precisely by defining the pair-wise interaction between participating atoms in the system. One of the novel ideas in this area could be using coarse-grained molecular dynamics to explore larger entities, such as grain structures, and study those coarse-grained morphologies in terms of their kinetic behavior to tackle the speed limitation of molecular dynamics.

\section*{Conflict of Interest Statement}

The authors declare that the research was conducted in the absence of any commercial or financial relationships that could be construed as a potential conflict of interest.

\section*{Author Contributions}

Mehrdad Yousefi conducted the research and prepared the draft of paper.

\section*{Funding}
No funding.

\section*{Acknowledgments}
No Acknowledgments.

\section*{Supplemental Data}
No Supplemental Data.

\section*{Data Availability Statement}
The datasets generated and analyzed for this study can be found in the [Continuum Field Grain Growth Github Repository.] [https://github.com/myousefi2016/Continuum-Field-Grain-Growth].



\clearpage

\section*{References}

\bibliographystyle{elsarticle-num}

\bibliography{md-ph}

\end{document}